# A micromagnetic study of the switching dynamics of the BiFeO₃/CoFe heterojunction

Yu-Ching Liao[1], Dmitri E. Nikonov[2], Sourav Dutta[3], Sou-Chi Chang[2], Chia-Sheng Hsu[1], Ian A. Young[2], and Azad Naeemi[1]

[1] School of Electrical and Computer Engineering, Georgia Institute of Technology, Atlanta, GA 30332 USA
[2] Components Research, Technology & Manufacturing Group Intel Corporation, Hillsboro, OR 97124 USA
[3] Department of Electrical Engineering, University of Notre Dame, Notre Dame, IN 46556, USA

**Abstract:**

The switching dynamics of a single-domain BiFeO₃/CoFe heterojunction is modeled and key parameters such as interface exchange coupling coefficient are extracted from experimental results. The lower limit of the magnetic order response time of CoFe in the BiFeO₃/CoFe heterojunction is theoretically quantified to be on to the order of 100 ps. Our results indicate that the switching behavior of CoFe in the BiFeO₃/CoFe heterojunction is dominated by the rotation of the Neel vector in BiFeO₃ rather than the unidirectional exchange bias at the interface. We also quantify the magnitude of the interface exchange coupling coefficient $J_{int}$ to be 0.32 pJ/m by comparing our simulation results with the giant magnetoresistance (GMR) curves and the magnetic hysteresis loop in the experiments. To the best of our knowledge, this is the first time that $J_{int}$ is extracted quantitatively from experiments. Furthermore, we demonstrate that the switching success rate and the thermal stability of the BiFeO₃/CoFe heterojunction can be improved by reducing the thickness of CoFe and increasing the length to width aspect ratio of the BiFeO₃/CoFe heterojunction. Our theoretical model provides a comprehensive framework to study the magnetoelectric properties and the manipulation of the magnetic order of CoFe in the BiFeO$_3$/CoFe heterojunction.

## I. Introduction:

Spintronic devices that utilize the spin degree of freedom of electrons to perform computation or store information are promising because of their non-volatility and high endurance[1]. To accomplish low power and magnetic-field free devices, researchers have pursued newly discovered physical phenomena such as the spin-transfer torque (STT), spin-orbit torque (SOT), and the magnetoelectric effect[2]. However, both STT and SOT effects require large current densities to switch magnets that result in large power dissipation and electromigration reliability issues caused by electromigration[3]. Magnetoelectric devices[4] on the other hand are voltage-controlled, which can alleviate the problem of a high threshold current and the accompanying joule-heating effect during and even after magnet switching. Current magnetoelectric devices utilize effects such as the voltage-controlled magnetic anisotropy (VCMA) effect in a composite material or the intrinsic magnetoelectric coupling in a single-phase multiferroic material. However, the VCMA effect that comes from the interface charge occupation levels can only rotate the magnetic easy-axis of the ferromagnet by $90^0$ by enhancing or reducing the perpendicular magnetic anisotropy energy. To flip the magnetic moment of a ferromagnet by $180^0$, which is needed for a magnetic tunnel junction, one has to apply either an extra spin current or a careful control of the applied voltage pulse width[5] for a deterministic magnet switching. Multiferroic materials that have at least two of the ferroic properties, including ferroelectricity, ferromagnetism and ferroelasticity, can successfully switch the magnetic moment of the adjacent ferromagnet by $180^0$ when a reverse electric field is applied, as demonstrated in the case of BiFeO$_3$ (BFO)[6,7].

BiFeO$_3$ is a multiferroic material with the properties of ferroelectricity, antiferromagnetism, and weak ferromagnetism at room temperature. The ferroelectricity of BFO originates from the movement of Bi$^{3+}$ ions under an applied electric field, and the saturation polarization in BFO is as large as 90 μC/cm$^2$ along [001] in a thin film structure[8]. In terms of magnetism, even though BFO is a G-type antiferromagnet, it has a weak magnetization that comes from Dzyaloshinskii-Moriya interaction[9,10] (DMI) because of the tilting of the oxygen octahedron. This weak magnetization in BFO is observed under a large magnetic field[11] or in a thin film structure[8] because the spin cycloid is destroyed by the large magnetic field or the broken symmetry from the epitaxial constraint in a thin film. Taking advantage of the coupling of the ferroelectric and antiferromagnetic orders in BFO and the interface exchange coupling between BFO and an adjacent CoFe layer, several experiments[7,12,13,14] have demonstrated voltage-controlled $180^0$ switching of the magnetic order in CoFe.

To evaluate the potential performance of this voltage-controlled BFO/CoFe heterojunction device in memory or logic devices, it is crucial to model the transient response, switching time, and the switching probability of the magnetic order in the CoFe layer. Nevertheless, previous studies have mainly focused on the domain patterns after switching[15], the shape anisotropy[16], the strain effect[17], and the strength of the interface exchange coupling field[13,14,15] in the multi-domain BFO/CoFe heterojunction. In this work, we develop a unified micromagnetic/ferroelectric simulation framework to model the switching dynamics and thermal stability of the single-domain BFO/CoFe heterojunction for the first time. In addition, other models[13,14,16] often consider the interface exchange field as an effective Zeeman field such that the mutual coupling of the magnetic orders in BFO and CoFe layers is neglected. In contrast, the interface exchange coupling field in our work is microscopically determined by Heisenberg exchange coupling and the weak exchange bias[18].

We analyze the interface properties of the BFO/CoFe heterojunction in section A. Then we quantitatively extract the magnitude of the interface exchange coupling coefficient $J_{int}$ based on the experimental data on giant magnetoresistance (GMR) curve[19] and magnetic hysteresis (M-H) loop measurement[7] as discussed in section B. Next, we demonstrate the switching dynamics of the ferroelectric BFO and the magnetic order of the BFO/CoFe heterojunction using the previous methodology[20] in section C. We also calculate the switching time limits of CoFe in a BFO/CoFe heterojunction by varying the ferroelectric polarization switching time in BFO. Last, we discuss the thermal stability and the switching probability of the BFO/CoFe device by analyzing the switching success rate and the energy barrier of the device in section D.

In our model, we consider a (001)-oriented BFO thin film grown on a $DyScO_3$ substrate with a CoFe thin film on top of the BFO. Our model of the BFO/CoFe heterojunction is set up according to the following steps. First, we consider the total energy terms in BFO and CoFe, as seen in the detailed descriptions in **Methods A**. BFO is a multiferroic material with ferroelectric energy, magnetic energy, and the magnetoelectric coupling energy which comes from the DMI. The coupling between BFO and CoFe is simulated by the interface exchange coupling. Next, the polarization dynamics and the spin dynamics of BFO after applying an electric field are solved by the Landau-Khalatnikov (LK) equation and the Landau–Lifshitz–Gilbert (LLG) equation, respectively. The simulation parameters of the ferroelectric model are listed in Table 1. The magnetic hard-axis of BFO, which is parallel to ***P***, will rotate simultaneously in the micromagnetic model following the experimental observation[21]. Overall, the rotation of ***P*** under an electric field would rotate both the antiferromagnetic order in BFO and the magnetization in CoFe because of the intrinsic magnetoelectric coupling in BFO and the BFO/CoFe interface exchange coupling, respectively. Our unified approach for simulating the ferroelectric and the magnetic dynamics simulation of the BFO/CoFe heterojunction is illustrated in Figure 1.

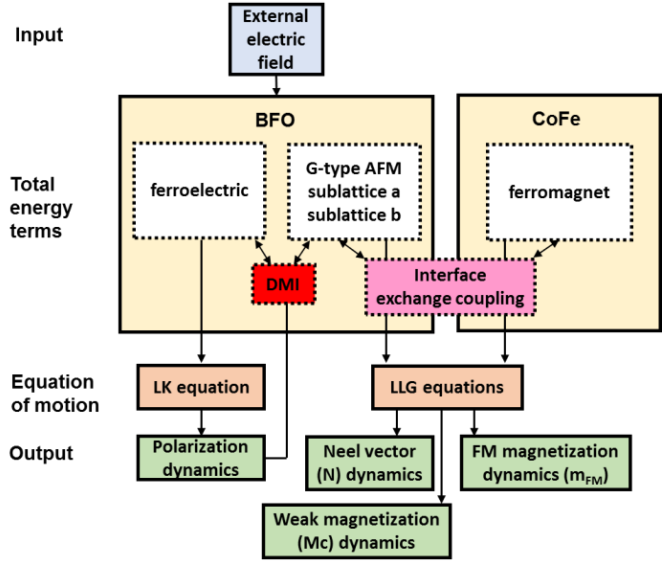

*Figure 1 Modeling framework for BiFeO₃/CoFe heterojunction.*

## II. Results and discussion

### A. Interface properties of the AFM/FM heterojunction

The interface exchange interaction and the magnetic order at the interface between BFO and CoFe thin films are investigated by the micromagnetic simulations (see detailed descriptions in **Methods B**). We define the Neel vector and the weak magnetization of BFO as $\boldsymbol{N} \equiv (\boldsymbol{M_1} - \boldsymbol{M_2})/(|\boldsymbol{M_1}| + |\boldsymbol{M_2}|)$ and $\boldsymbol{M_c} \equiv (\boldsymbol{M_1} + \boldsymbol{M_2})/(|\boldsymbol{M_1}| + |\boldsymbol{M_2}|)$, respectively, where $\boldsymbol{M_1}$ and $\boldsymbol{M_2}$ are the magnetization vectors of sublattices *1* and *2* in BFO. The magnetization of CoFe is expressed as $\boldsymbol{M_{FM}}$. In the steady state before an electric field is applied, the weak magnetization in BFO shows a staggered behavior when averaged in the *z* direction (out-of-plane direction), as shown in the inset of Figure 2 (a). The thickness of BFO is 30 nm, which corresponds to *z*=30 nm in Figure 2 (a). The staggered $M_c$ arises from the antiferromagnetic exchange coupling inside BFO. Besides, when $M_c$ is close to the interface of CoFe, the magnitude of $M_c$ is stronger than that of $M_c$ in the bulk region of BFO due to the strong interface exchange coupling between BFO and CoFe layers. Therefore, by comparing the staggered $M_c$ under weak or strong DM fields ($H_{DM}$) in the bulk BFO, i.e. $H_{DM} = 0\ Oe$ versus $H_{DM} = 10^4\ Oe$ in Figure 2 (a) and (b), we find that when DM field is strong, there is a finite $M_c$ in the bulk region because of the stronger asymmetric DM field. However, when the DM field is weak, $M_c$ in the bulk region is nearly zero since the antiferromagnetic exchange coupling field is stronger than the DM field. In addition to the magnitude of DM field, the magnitude of $M_c$ ($|M_c|$) also depend on the tilting of the magnetic moments at the BFO/CoFe interface as seen in Figure 2 (a) and (c).

Previous studies[7,15] have also shown that $|M_c|$ in the BFO layer of a BFO/CoFe heterojunction is enhanced compared to that of $|M_c|$ in a stand-alone BFO layer. To check the consistency of our model with the experimental observations from previous studies, we consider two scenarios with various magnitudes of $H_{DM}$ in BiFeO3: a stand-alone 30 nm thick BFO layer versus a 30 nm thick BFO/2 nm thick CoFe heterojunction. Figure 2 (d) shows that the enhancement of $|M_c|$ is larger at low $H_{DM}$ than at high $H_{DM}$ because the interface exchange coupling field is relatively stronger than the bulk DMI at low $H_{DM}$. Furthermore, we observe that under high $H_{DM}$, $|M_c|$ of a stand-alone BFO coincides with the case of BFO/CoFe, which indicates that $M_c$ becomes dominated by the bulk DMI rather than the interface exchange field. Figure 2 (d) also shows that $|M_c|$ at $z$ =30 nm is stronger than $|M_c|$ at $z$ =28 nm, and these interfacial $|M_c|$ values are insensitive to the magnitude of $H_{DM}$ when $H_{DM}$ is smaller than $10^4$ Oe. Therefore, the behavior of the interface properties, the interface exchange coupling, and $|M_c|$ all depend on the value of $J_{int}$. On one hand, $J_{int}$ is desired to be large enough to provide strong coupling between the antiferromagnet and ferromagnet. On the other hand, $J_{int}$ needs to be smaller than the intrinsic exchange coupling inside BFO and CoFe.

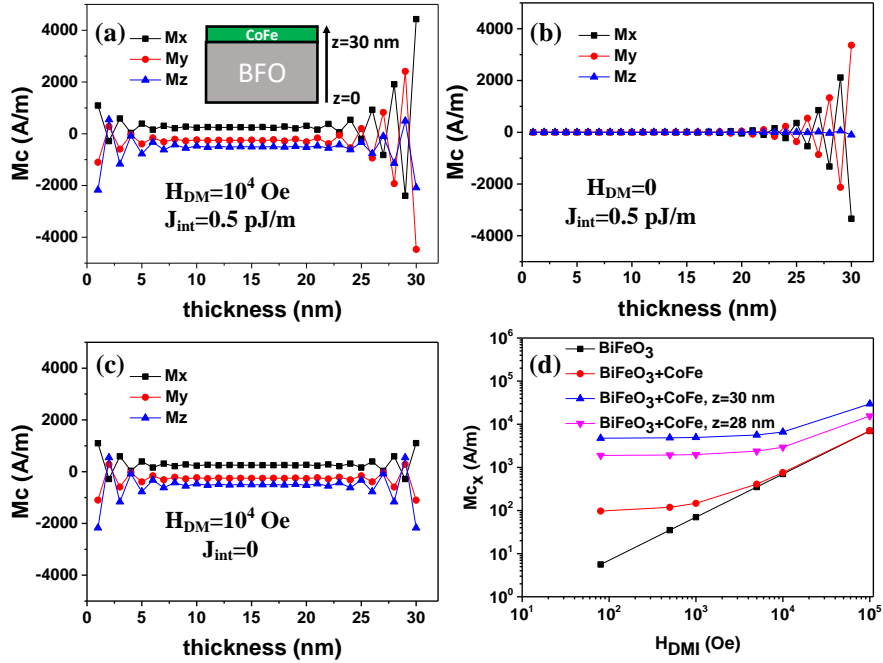

Figure 2 (a) $M_c$ in 32 nm BFO with $H_{DMI} =10^4$ Oe, $J_{int}=0.5$ pJ/m. (b) $M_c$ in 32 nm BFO with $H_{DMI}=0$ Oe, $J_{int}=0.5$ pJ/m. (c) $M_c$ in 32 nm BFO with $H_{DMI} =10^4$ Oe, $J_{int}=0$ pJ/m.. (d) The magnitude of $M_c$ in the BFO(black) or in the BFO/CoFe heterojunction at the interface (blue), close to the interface(magenta), or the average value(red) when $J_{int}=0.5$ pJ/m.

### B. Experimental verification of interface properties

#### a. GMR

To extract the simulation parameter $J_{int}$ from experimental data, we consider a 30 nm thick BFO thin film stack with 2 nm thick CoFe thin film (free layer ferromagnet (FM)), 2 nm thick Cu, and 2 nm thick top layer CoFe (reference layer FM) with $L$ = 2 μm and $W$ = 200 nm , as shown in the structure 1 of Figure 3 (a). The thickness of BFO is chosen to be thinner than the actual thickness in the experiment (~100 nm) to save the computation time. Note that the BFO thin film has a stripe domain pattern with a domain width of about 200 nm, as observed in the experiments[19].

Next, to compare with the GMR in the experiment[19], we sweep the external magnetic field from -300 Oe to 300 Oe with varying magnitudes of $J_{int}$. The resistance of the spin-valve during the GMR measurement is determined by

$$R = R_0 + \Delta R \sin^2\frac{\theta}{2}, \tag{1}$$

where $R_0$ is the resistance of the ferromagnet sublattices, and $\theta$ is the angle between the magnetizations of the two CoFe layers. The GMR ratio is then calculated by

$$GMR = (R - R_P)/R_P = \Delta R \sin^2(\theta/2)/R_0, \tag{2}$$

where $R_P$ is the resistance when the bottom and top layers of CoFe are parallel. We find that the curve of GMR ratio is highly dependent on the magnitude of $J_{int}$, as shown in Figure 3 (c). This is because when $J_{int}$ increases, the magnetic coercive field ($H_c$) of the free layer FM increases due to the strong interface exchange coupling to the BFO layer (as seen in Figure 3 (e)); hence, the curve of GMR ratio broadens. Similarly, when $J_{int}$ is small, the free layer FM becomes decoupled from the BFO layer; as a consequence, $H_c$ decreases and the GMR curve becomes narrowly peaked. By varying the magnitude of $J_{int}$, we find that when $J_{int}$ is equal to 0.32 pJ/m, our model fits well with the experimental data[19] without considering any non-ideality effects such as nucleation of defects. In addition, we compare the curve of the GMR ratio when the aspect ratio of BFO spin-valve structure changes from 100 to 2. Figure 3 (d) presents simulation results that are similar to the experimental data in which the distance between two peaks in the GMR curve decreases because of a smaller $H_c$ in the reference layer FM. There is also no obvious exchange bias in the hysteresis loop of the free layer FM because of the larger width of the BFO layer. Consequently, the exchange bias of different domains will be compensated, as shown in Figure 3 (f).

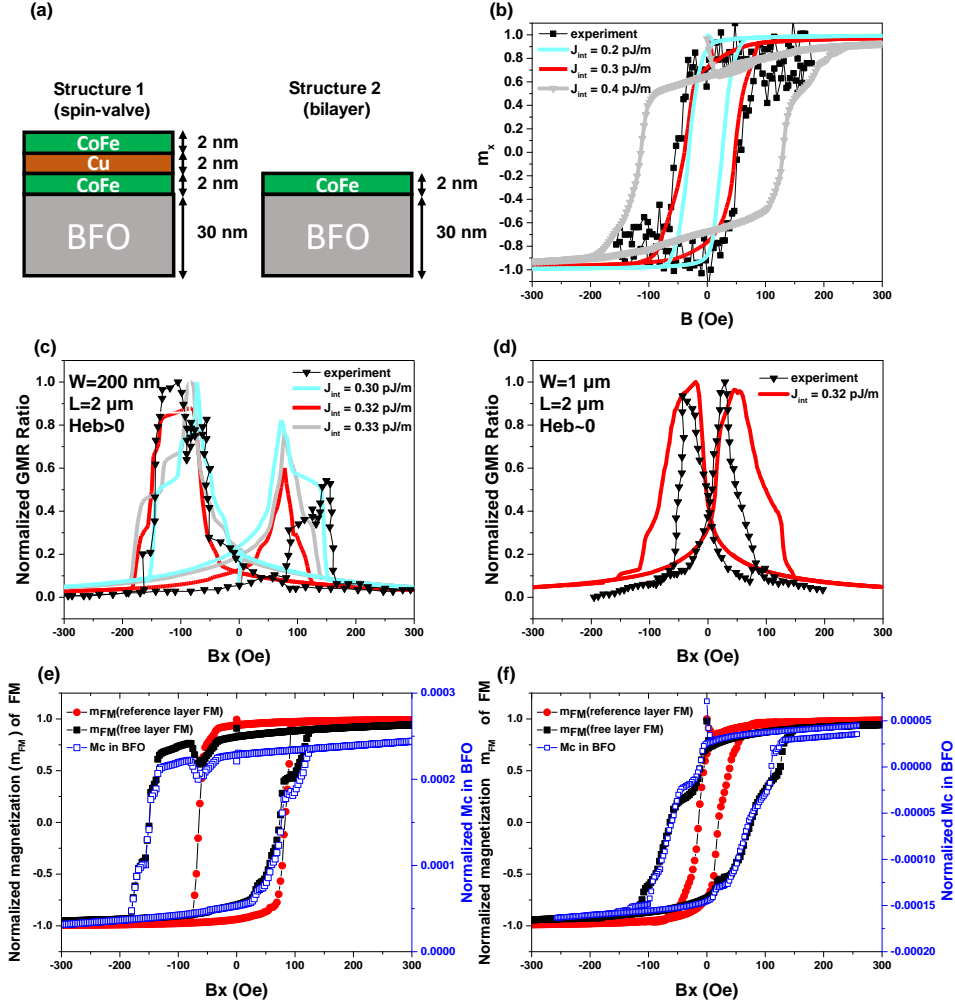

*Figure 3 (a) Schematics of the device cross section in the GMR (structure 1) and the magnetic hysteresis loop (structure 2) simulations. Comparison between fitted GMR curve when (c) W=200 nm, L=2 μm and (d) W=1 μm, L=2 μm and experiment[19] for $J_{int}$ extraction. The magnetic hysteresis loops of BFO, free layer ferromagnet, and reference layer ferromagnet when $J_{int}$ = 0.32 pJ/m with W =200 nm (e) or W=1 μm (f). (b) Comparison between fitted magnetic hysteresis loop (M-H loop) and experiment[7].*

### b. M-H loop

To confirm the magnitude of $J_{int}$ extracted in sub-section a, we use another set of experimental M-H loop measurements[7]. We simulate a stripe-domain 30 nm thick BFO and 2 nm thick CoFe bilayer thin films, as shown in structure 2 in Figure 3 (a). Similar to the experiments[7], the lateral dimensions are 5 μm long and 2 μm wide. We sweep the external magnetic field from - 300 to 300 Oe and check the M-H loop of the BFO/CoFe heterojunction. Figure 3 (b) shows that our simulation results using $J_{int}= 0.3$ pJ/m is in a good agreement with the experimental data. This result further confirms the magnitude of $J_{int}$ is close to 0.32 pJ/m when the mesh size for numerical simulations is $20\times20\times2$ nm$^3$.

### C. Switching dynamics of BFO/CoFe bilayer thin film

#### a. Switching dynamics of BFO/CoFe bilayer thin film

Next, we investigate the dynamics of a single-domain 30 nm thick BFO and 2 nm thick CoFe bilayer thin films. The width (W) and length (L) are considered to be 20 nm. We renormalize the magnetic parameters of BFO and $J_{int}$ with smaller mesh size $5\times5\times1$ nm$^3$ to simulate a scaled device. The details of the renormalization approach are explained in **Method B**. Theoretically, the dynamics of CoFe is determined by the magnetization rotation in BFO because of the interface exchange coupling between $M_c$ and $M_{FM}$. To simulate the dynamics of both BFO and CoFe, we use the modeling framework for a single-domain BFO from our previous study[20], in which the dynamics of $P$, $N$, and $M_c$ are calculated by considering the rotation of the magnetic easy-plane along with the polarization[21]. Similar to the case of the single-domain BFO, we see that $N$ switches $180^0$ whereas $M_c$ does not, as shown in Figure 4 (a). In addition, $M_{FM}$ switches $180^0$ after the polarization reversal, and there exists an intermediate stage when $M_{FM}$ rotates $90^0$ because of the two-step polarization switching characteristics of BFO[7]. This result is interesting since it has been argued before that the magnetization reversal in CoFe mainly comes from the exchange coupling field between $M_c$ and $M_{FM}$[14]. However, our results indicate that the main driving force is the rotation of $N$ rather than the rotation of $M_c$ since $|M_c|$ is much smaller than $|N|$. To verify the statement that the rotation of $N$ governs the switching behavior of $M_{FM}$, we consider two BiFeO$_3$ switching scenarios: (i) the magnetic easy-plane rotates with the magnetic hard-axis parallel to $P$, and (ii) the magnetic easy-axis of BiFeO$_3$ is fixed, but the DM field switches $180^0$ in the direction of $\boldsymbol{P} \times \boldsymbol{N}$. In case (i), Figure 4(a) shows that when $\boldsymbol{N}$ switches $180^0$, $\boldsymbol{M_{FM}}$ successfully rotates $180^0$ after polarization reversal whereas $\boldsymbol{M_c}$ remains unswitched. On the contrary, in case (ii), when $\boldsymbol{N}$ is fixed during polarization reversal, $\boldsymbol{M_c}$ switches but neither $\boldsymbol{M_{FM}}$ nor $\boldsymbol{N}$ switches, as shown in Figure 4 (b). Therefore, we believe that the magnetic switching of $\boldsymbol{M_{FM}}$ is driven by the rotation of $\boldsymbol{N}$ rather than $\boldsymbol{M_c}$.

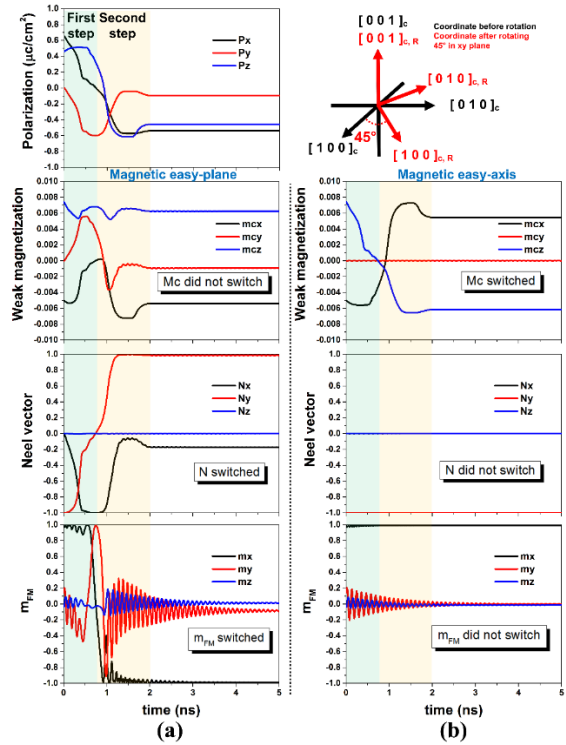

*Figure 4 The dynamics of the polarization ($\boldsymbol{P}$), the weak magnetization ($\boldsymbol{M_c}$), the Neel vector ($\boldsymbol{N}$), and the magnetization of ferromagnet (CoFe) in the BiFeO$_3$/CoFe heterojunction when the magnetic state of BFO is an (a) easy-plane state or (b) easy-axis state.*

b. **Sensitivity analysis of the magnet switching probability with varying polarization switching time ($t_{FE}$)**

Next, to evaluate the potential application of the BFO/CoFe heterojunction in logic/memory operations, we check the sensitivity of BFO polarization switching time ($t_{FE}$) that may affect the magnetization switching success rate and the magnetization switching time of CoFe. Note that varying $t_{FE}$ of the BFO corresponds to various viscosity coefficients in the LK equation under a fixed electric field.

In our previous study[20], we have shown that the theoretical limit of the switching time of **N** in BFO is 30 ps; therefore, the antiferromagnetic order of BFO can follow the rotation of the polarization. For a BFO/CoFe heterojunction with $L = 40$ nm, $W = 20$ nm, and 30 nm thick BFO and 2 nm thick CoFe, Figure 5 (a) demonstrates that when $t_{FE}$ is shorter than 1.45 ns, the rotation of **N** is too fast for $\boldsymbol{M_{FM}}$ to follow; thus, the switching of $\boldsymbol{M_{FM}}$ failed. In fact, we find that the switching time of $\boldsymbol{M_{FM}}$ follows the input switching time of **N**; in other words, the increasing $t_{FE}$ increases the switching time of **N**, $\boldsymbol{M_c}$ and thus $\boldsymbol{M_{FM}}$. Therefore, Figure 5 (a) shows that the theoretical lower bound for $\boldsymbol{M_{FM}}$ switching is 1.45 ns when $L=40$ nm, $W=20$ nm, and CoFe is 2 nm thick. However, this minimum $t_{FE}$ for successful magnet switching ($t_{FE,min}$) of the BFO/CoFe heterojunction may not be a constant value since $t_{FE,min}$ depends on the strength of the interface exchange coupling between BFO and CoFe.

To examine the $t_{FE,min}$ dependency on different energy barriers in the BFO/CoFe heterojunction or different strength of the interface exchange coupling in CoFe, we consider two scenarios: (A) varying aspect ratios of the device and (B) varying thicknesses of the CoFe under a fixed width (20 nm) of the BFO/CoFe heterojunction and a fixed thickness of BFO (30 nm). For case (A), Figure 5 (b-d) show the relation between $t_{FE,min}$ and the varying aspect ratios of the BFO/CoFe bilayer. The minimum $t_{FE}$ for successful magnet switching increases as the aspect ratio of the device increases because the larger energy barrier requires the longer response time in CoFe under a constant interface exchange coupling field. In other words, the overdrive field for the CoFe layer decreases if the aspect ratio of the device increased. However, $t_{FE,min}$ will become less dependent on the aspect ratio when the thickness of the CoFe decreases below 1 nm. This is because the interface exchange coupling field becomes dominant compared to the intrinsic exchange coupling field in thin CoFe. Therefore, for case (B), when the thickness of the CoFe is varied from 0.5 nm to 2 nm, $t_{FE,min}$ monotonically increases because of the weaker interface exchange coupling averaged in the CoFe, as shown in Figure 5 (e). These time-dependent magnetization switching cases show that $t_{FE,min}$ in the BFO/CoFe heterojunction is 90 ps when the CoFe is as thin as 0.5 nm, and the effect of the interface exchange coupling field degrades with increasing CoFe thickness. To be more specific, when the thickness of CoFe increases from 1.5 nm to 2 nm, $t_{FE,min}$ drastically increases from 230 ps to 660 ps.

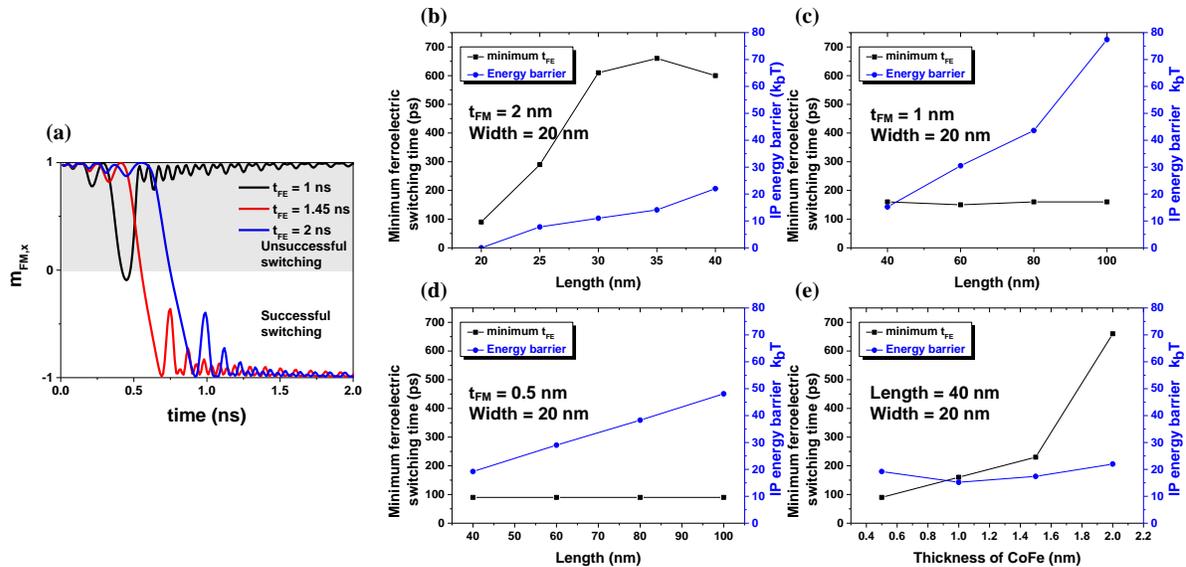

*Figure 5 (a)The minimum ferroelectric switching time ($t_{FE}$) for successful magnet switching in the BFO/CoFe heterojunction when length is 40 nm, width is 20 nm, and the thickness of BFO and CoFe are 30 nm and 2 nm, respectively. The minimum ferroelectric switching time of CoFe with varying lengths of the device when the thickness of CoFe is (b) 2nm, (c) 1 nm, and (d) 0.5 nm. (e) The minimum ferroelectric switching time of CoFe with varying thicknesses of CoFe when length is 40 nm and width is 20 nm.*

### D. Thermal stability of the BFO/CoFe bilayer thin film

#### a. Energy landscape

To study the thermal stability of the device, we check the magnetic energy barrier of the 20 nm wide single-domain BFO/CoFe thin film with varying device length and CoFe thickness. Since our calculations show that the out-of-plane (OOP) energy barrier is much larger than the in-plane (IP) energy barrier due to the strong shape anisotropy energy, we will only discuss the IP energy barrier in the following section. When the thickness of CoFe is 2 nm, the IP energy barrier increases from 0 to 23 $k_bT$ when the length increases from 20 nm to 40 nm, as shown in Figure 5 (b). These energy barriers are too small for memory applications which usually require energy barriers larger than 60 or 70 $k_bT$ for retention times larger than ten years. When the thickness of CoFe is 1 nm, the energy barrier of the BFO/CoFe bilayer is as large as 77.4 $k_bT$ in the 100 nm long BFO/CoFe bilayer, as seen in Figure 5 (c). However, when the thickness of CoFe is reduced to 0.5 nm in the 100 nm long BFO/CoFe bilayer, the IP energy barrier decreases to 48 $k_bT$ because of the reduced volume of CoFe. Therefore, to ensure a high IP energy barrier, a 1 nm thick FM layer and a high length to width aspect ratio is suitable for the BFO/CoFe heterojunction.

#### b. Probability of successful switching

The switching dynamics of the BFO/CoFe bilayer under the thermal noise effect is checked by using the stochastic Landau–Lifshitz–Gilbert (LLG) equation:

$$\dot{\boldsymbol{M}} = -\gamma(\boldsymbol{M} \times \boldsymbol{H}) + \frac{\alpha_G}{M_s}(\boldsymbol{M} \times \dot{\boldsymbol{M}}) \qquad (3)$$

where the effective field $\boldsymbol{H}$ includes the Gaussian white noise term $\boldsymbol{H_T}(t)$. The thermal field $\boldsymbol{H_T}(t)$ obeys the properties of Brownian motion, meaning that the mean $\langle \boldsymbol{H_T}(t) \rangle = \boldsymbol{0}$ and the correlation function over a time interval $\tau$ satisfies

$$\langle \boldsymbol{H_T}(t)\boldsymbol{H_T}(t+\tau) \rangle = \frac{2k_bT\alpha}{\gamma\mu_0^2 M_s V}\delta(\tau). \qquad (4)$$

Note that this stochastic thermal field varies with time and may affect the mean first passage time (MFPT), i.e. the time when the magnetic states in one energy minimum $E_0$ flip to another energy minimum, as seen in Figure 6 (a). The positive and negative exchange bias of the BFO/CoFe heterojunction are represented by the two energy minima which determine the magnetization direction of CoFe. Generally, the MFPT depends on the energy barrier of the device, the second order derivative of the minimum potential, and the temperature [22]. This MFPT increases with the higher energy barrier and the lower temperature.

To investigate the thermal stability and switching reliability of the device, we check the probability of successful switching in the BFO/CoFe heterojunction with varying aspect ratios and thicknesses of CoFe. The probability of successful switching ($P_{sw}$) is obtained by simulating the polarization and the magnetization switching dynamics of the BFO/CoFe heterojunction for 20 tests. Figure 6 (b) shows that, for the 30 nm thick BFO/2 nm thick CoFe heterojunction, the switching of CoFe succeeds when the length is 40 nm but fails when lengths longer than 45 nm because the IP energy barrier becomes larger than the interface exchange coupling energy as the aspect ratio increases. However, $P_{sw}$ is only 50 % in the 20 tests when CoFe is 2 nm thick and $L$=40 nm, which means that the device of this size is not ideal for real applications. For the 1 nm thick CoFe with $L$=100 nm, we observe that $P_{sw}$ is 100% in these 20 tests because of the stronger interface exchange coupling effect between BFO and thin CoFe layer.

From the analysis of the $t_{FE,min}$, IP energy barrier and $P_{sw}$, our results show that the thermal stability and the magnet switching time of the BFO/CoFe heterojunction depend on the proper design of the aspect ratio and the thickness of CoFe. Generally, both a faster magnetization switching (smaller $t_{FE,min}$) and a high switching success rate are obtained from thinner CoFe films. On the other hand, when the aspect ratio is 5, the energy barrier will greatly decrease as the thickness of CoFe reduces from 1 nm to 0.5 nm since the energy barrier depends on the volume of the ferromagnet. Therefore, a BFO/CoFe heterojunction with a 1 nm thick CoFe and a high aspect ratio seems to be the most promising option in terms of thermally stability and error free operation.

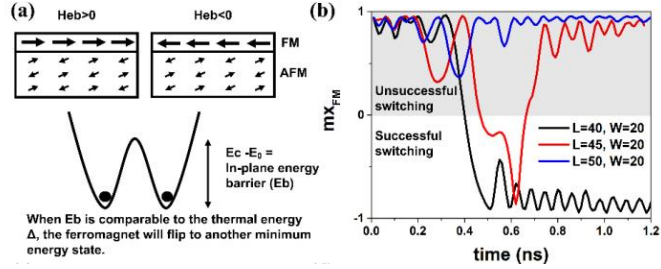

*Figure 6 (a) The schematic of the magnetic states under different polarity of the exchange bias. The two stable states of the AFM/FM heterojunction correspond to two energy minima in the double-well energy profile. When thermal noise is included in the simulation, the system may flip from one stable state to another state depending on the energy barrier of the system and the curvature of the two energy minima in the energy profile. (b) Comparison of the normalized magnetization switching curves with varying length of the BFO/CoFe heterojunction under fixed width.*

### III. Conclusion

We have modeled the magnetic structure and the dynamics of a single domain BiFeO$_3$/CoFe heterojunction. In the equilibrium state, we show that $M_c$ in BFO is strongly enhanced by $M_{FM}$ and $J_{int}$ through the interface exchange coupling with the CoFe layer. Moreover, $M_c$ has a staggered behavior close to the interface of CoFe even though bulk DMI is negligible because of the dominant interface exchange coupling effect. To evaluate the interface properties and determine the value of $J_{int}$, we model the experimentally measured GMR and magnetic hysteresis loop, and find that $J_{int}$ is approximately $0.32 \ pJ/m$. Using this experimentally extracted $J_{int}$, we further simulate the dynamics of BFO/CoFe heterojunction and prove that the driving force of $M_{FM}$ switching is determined by the rotation of $N$ rather than the rotation of $M_c$. We also analyze the sensitivity of the switching success rate of the BFO/CoFe heterojunction with varying polarization switching time ($t_{FE}$). The minimum $t_{FE}$ ($t_{FE,min}$) of a successful switching depends on the aspect ratio of the device and the thickness of the CoFe layer. It is found that a smaller $t_{FE,min}$ can be obtained by a thinner CoFe thin film. Last, we include the thermal noise effect of this BFO/CoFe heterojunction to evaluate the thermal stability and the probability of successful switching for its further applications at room temperature. Our simulation results show that for a BFO/CoFe heterojunction with a thinner FM thin film (~1 nm), we can increase the aspect ratio to about 5 to ensure both high thermal stability and switching reliability ($P_{sw}$~100%). The results of this paper are important for understanding and designing magnetoelectric devices.

## IV. Methods

### A. Total energy terms in BFO and CoFe

The total energy in BFO includes the ferroelectric energy, magnetic energy, and the magnetoelectric coupling energy. The ferroelectric energy of BFO includes the bulk energy $F_{bulk}(\boldsymbol{P})$, elastic energy $F_{elas}(\boldsymbol{P})$ and the electric energy $F_{elec}(\boldsymbol{P})$ which are expressed as

$$F_{bulk}(\boldsymbol{P}) = \alpha_1(\boldsymbol{P}_1^2 + \boldsymbol{P}_2^2 + \boldsymbol{P}_3^2) + \alpha_{11}(\boldsymbol{P}_1^4 + \boldsymbol{P}_2^4 + \boldsymbol{P}_3^4) + \alpha_{12}(\boldsymbol{P}_1^2\boldsymbol{P}_2^2 + \boldsymbol{P}_1^2\boldsymbol{P}_3^2 + \boldsymbol{P}_2^2\boldsymbol{P}_3^2),$$

$$F_{elas}(\boldsymbol{P}) = K_{Strain}(\boldsymbol{P} \cdot \boldsymbol{u})^2,$$

$$F_{elec}(\boldsymbol{P}) = -\left(c\frac{(\boldsymbol{P}_{dw} - \boldsymbol{P}) \cdot \boldsymbol{P}}{\epsilon_r \epsilon_0} + \boldsymbol{P} \cdot \boldsymbol{E}_{ext}\right),$$

where $\alpha_1, \alpha_{11}, \alpha_{12}$ are the phenomenological Landau expansion coefficients, $\boldsymbol{u}$ is the axis of strain, $K_{Strain}$ is the strain energy, $c$ is the geometry factor of the averaged domain wall, $\boldsymbol{P}_1, \boldsymbol{P}_2$, and $\boldsymbol{P}_3$ are the polarization components in x, y, and z- directions, and $\boldsymbol{E}_{ext}$ is the external electric field. The magnetic energy of BFO in cell $i$ includes the exchange energy within BFO $F_{exchange,BFO}(\boldsymbol{M}_{i,AFM})$, the interlayer exchange coupling between BFO and CoFe $F_{exchange,int}(\boldsymbol{M}_{i,AFM}, \boldsymbol{M}_{i,FM})$, the bulk anisotropy energy $F_{ani,bulk}(\boldsymbol{M}_{i,AFM})$, the anisotropy energy that comes from the epitaxial constraint $F_{ani,epi}(M_{i,AFM})$, and the demagnetization energy $F_{dem}(M_{i,AFM})$. The magnetic energy of each term in BFO is expressed as:

$$F_{exchange,BFO}(M_{i,AFM}) = J_{AFM}(\vec{\nabla} m_{i,AFM})^2,$$

$$F_{exchange,int}(M_{i,AFM}, M_{i,FM}) = J_{int} \frac{m_{i,AFM} \cdot (m_{i,AFM} - m_{i,FM})}{\Delta_{ij}^2},$$

$$F_{ani,bulk}(M_{i,AFM}) = K_{AFM}(m_{i,AFM} \cdot \widehat{P})^2,$$

$$F_{ani,epi}(M_{i,AFM}) = K_{epi} m_{z,AFM}{}^2,$$

$$F_{dem}(M_{i,AFM}) = -\frac{1}{2}\mu_0 M_s (H_{dem,i} \cdot m_{i,AFM}),$$

where $J_{AFM}$ is the exchange constant of BFO, $J_{int}$ is the interlayer exchange coupling coefficient between BFO and CoFe, $m_{i,AFM} = M_{i,AFM}/M_{s,AFM}$ is the unit magnetization vector of BFO, $m_{i,FM} = M_{i,FM}/M_{s,FM}$ is the unit magnetization vector of CoFe, $M_{s,AFM}$ and $M_{s,FM}$ are the saturation magnetization of the BFO and CoFe, respectively, $K_{AFM}$ is the bulk anisotropy energy that comes from DM interaction, $\widehat{P}$ is the hard axis of the antiferromagnet, $K_{epi}$ is the anisotropy energy which is along [001], and $H_{dem,i}$ is the demagnetization field of cell $i$. The magnetoelectric energy of BFO is expressed as

$$F_{me}(P, M_i) = F_{DMI} = \sum_{i=1}^{N} D_{i,j} \cdot (N \times M_c),$$

where $D_{i,j} = V_0(r_{ij} \times x) \approx V_0 \widehat{P}$ is the DM vector with the direction parallel to the polarization of the BFO and $V_0$ is the magnitude of the DMI energy.

The total energy of CoFe includes the exchange energy within CoFe, the interlayer exchange coupling energy at the BFO/CoFe interface, and the demagnetization energy. The magnetic energy of each term in CoFe is expressed as:

$$F_{exchange,CoFe}(M_{i,FM}) = J_{FM}(\vec{\nabla} m_{i,FM})^2,$$

$$F_{exchange,int}(M_{i,AFM}, M_{i,FM}) = J_{int}\frac{m_{i,FM} \cdot (m_{i,FM} - m_{i,AFM})}{\Delta_{ij}^2},$$

$$F_{dem}(M_{i,FM}) = -\frac{1}{2}\mu_0 M_s (H_{dem,i} \cdot m_{i,FM}),$$

where $J_{FM}$ is the exchange constant of CoFe. Note that the bulk anisotropy energy of CoFe is neglected since CoFe with thin thickness is amorphous.

### B. Micromagnetic simulation

In the micromagnetic model, we use the Object Oriented MicroMagnetic Framework (OOMMF)[23] to implement the modeling of both BFO and CoFe layers. Since BFO is a G-typed antiferromagnet with a weak-canted magnetization generated by DMI, we incorporate two sublattices *1* and *2*, and define two unitless vectors $N$ as Neel vector and $M_c$ as the weak-canted magnetization where $N \equiv (M_1 - M_2)/(|M_1| + |M_2|)$ and $M_c \equiv (M_1 + M_2)/(|M_1| + |M_2|)$.

The total effective magnetic field of the magnetic structure in BFO is calculated from the derivative of the total energy with respect to the magnetization. The total effective magnetic field includes the exchange coupling field $H_{ex}$, the bulk anisotropy field $H_{ani}$ from DMI, and the anisotropy field $H_{epi}$ from compressive epitaxial constraint, the demagnetization field $H_{dem}$, and the unidirectional DMI field $H_{DMI}$ which approximately equals a Zeeman field that creates spin canting in BiFeO$_3$. The direction of $H_{DMI}$ is determined by $N \times D$ since the Hamiltonian of DMI is expressed as $E_{DMI} = -D \cdot \sum M_i \times M_j = -D \cdot \sum N \times M_c = \sum H_{DMI} \cdot M_c$, where $H_{DMI} = N \times D$, $|D| = \beta P$ and $\beta$ is the magnetoelectric constant[10]. The Hamiltonian of DMI also shows a right-handed relation between the direction of $P, N,$ and $M_c$[24]. The single-ion anisotropy (SIA) energy of BiFeO$_3$ is neglected since experiments[25] proved that the magnitude of the SIA energy is 20 times smaller than $H_{ani}$. For the magnetoelectric coupling in BiFeO$_3$, the hard axis of the bulk anisotropy field in BiFeO$_3$ is parallel or antiparallel to the direction of $P$ according to the experimental observation from previous studies[21,26].

In CoFe thin film, the total effective magnetic field includes the exchange coupling field $H_{ex}$, the interface exchange coupling field $H_{int}$ of the nearest neighbors, and the demagnetization field $H_{dem}$. The interface exchange coupling field is numerically expressed as $H_{int,ij} = \frac{2J_{int}}{\mu_0 M_{s,FM}} \sum \frac{m_{FM}(r_i+\Delta_j) - m_{AFM}(r_i)}{|\Delta_j^2|}$, where $m_{AFM}(r_i)$ is the magnetization of BiFeO$_3$ in the cell $i$, $\Delta_j \in \Delta_z$ is distance from the cell $i$ to the nearest neighbor cell $j$ in $z$ direction, and $m_{FM}(r_i + \Delta_j)$ is the magnetization of the nearest neighbor cell $j$ in CoFe with distance $\Delta_j$. The growth anisotropy and crystalline anisotropy energy of CoFe are neglected since the thickness of the CoFe thin film is only 0.5 to 2 nm in a polycrystalline structure.

The dynamics of both BFO and CoFe are solved by using the LLG equation as Equation (3). The simulation parameters of the micromagnetic model are summarized in Table 2 and Table 3.

**B. Renormalization of magnetic parameters in BFO and the $J_{int}$**

To simulate a G-type antiferromagnet (AFM) in OOMMF, the exchange stiffness coefficient of AFM needs to be normalized with varying mesh sizes and shapes.

The exchange energy in OOMMF is expressed as $E_i = \sum_{j \in N_i} A_{ij} \frac{m_i \cdot (m_i - m_j)}{\Delta_{ij}^2}$, where $N_i$ is the set of the 6 nearest cells, $A_{ij}$ is the exchange coefficient between cell $i$ and cell $j$, and $\Delta_{ij}$ is the discretization steps between the cell $i$ and the cell $j$. This exchange energy comes from the approximation of a Heisenberg-type exchange coupling assuming the magnetization $m_i$ is continuous. This means that the total exchange energy $E_{exch} = -2J\sum_{i<j}^{nn}J_{ij}S_iS_j \cong -2J\sum_{i<j}^{nn}S_iS_j = -2\sum_{i<j}J_{ij}S^2\cos\theta_{ij}$ where $J$ is the exchange coupling constant, $S_i, S_j$ are the spin vectors of sublattice $i$ and $j$, and $\theta_{ij}$ is the angle between vectors $S_i$ and $S_j$. Here the factor of two denotes that there is exchange coupling energy from cell $i$ to cell $j$ and vice versa. Suppose $\theta_{ij}$ is small, we can approximate $E_{exch} \cong -2\sum_{i<j}J_{ij}S^2(1-\frac{1}{2!}\theta_{ij}^2) = \sum_{i<j}J_{ij}S^2\theta_{ij}^2 + constant$ by using the Taylor expansion. Since $\theta_{ij}$ is a continuous variable, $\theta_{ij}$ can be approximated as $\theta_{ij} \cong a\frac{\partial\theta}{\partial x}$ in the numerical simulation where $a$ is the discretization step. However, this assumption that $\theta_{ij}$ is small is invalid for an AFM material since the magnetic moment in an AFM is staggered in a positive to negative direction. Therefore, the exchange energy is non-convergent for AFM when using the normal numerical expression of the exchange coupling as used in a FM.

To simulate the exchange coupling field in AFM directly from the spin Hamiltonian, we have done a micromagnetic simulation in MATLAB using $H_{exch}^{ij} = -\frac{1}{\mu_0 M_s a^3}\frac{\partial E_{exch}^{ij}}{\partial \alpha_j} = \frac{JS^2}{\mu_0 M_s a^3}(\alpha_i^x, \alpha_i^y, \alpha_i^z) = \frac{2A}{\mu_0 M_s a^2}(\alpha_i^x, \alpha_i^y, \alpha_i^z)$, where $\alpha_j$ is the direction cosines of the cell $j$, $\alpha_i$ is the direction cosines of the neighboring cell $i$, $a$ is the lattice constant, $E_{exch}^{ij} = -JS^2(\alpha_j^x\alpha_i^x + \alpha_j^y\alpha_i^y + \alpha_j^z\alpha_i^z)$, and $A = \frac{JS^2}{2a} < 0$. Our results show the same result as done in OOMMF when using the same mesh size, i.e. 1×1×1 nm³ as shown in Figure S 1. However, the simulation results will vary when we change the mesh size in OOMMF since the numerical expression of the exchange coupling field is nonconvergent for an AFM. To calculate the exchange coupling field in an AFM layer in OOMMF using larger mesh sizes under the same exchange energy density, the exchange stiffness constant needs to be renormalized. The renormalization of the exchange stiffness constant is done by considering the number of nearest neighbors in each cell with varying mesh sizes. The exchange energy in AFM is expressed numerically as

$$E_{ex,AFM} = J_{AFM}\left(\left[\sum_{i\in 6\ n.n.}\frac{m_i(r_i)\cdot(m_i(r_i)-m_i(r_i\pm\Delta_j))}{\Delta_{ij}^2} + \sum_{i\in 5\ n.n.}\frac{m_i(r_i)\cdot(m_i(r_i)-m_i(r_i\pm\Delta_j))}{\Delta_{ij}^2} + \sum_{i\in 4\ n.n.}\frac{m_i(r_i)\cdot(m_i(r_i)-m_i(r_i\pm\Delta_j))}{\Delta_{ij}^2} + \sum_{i\in 3\ n.n.}\frac{m_i(r_i)\cdot(m_i(r_i)-m_i(r_i\pm\Delta_j))}{\Delta_{ij}^2}\right]\right),$$

where $J_{AFM}$ is the exchange stiffness constant of AFM, $n.n.$ denotes the nearest neighbors of the cell $i$ and $\Delta_{ij}$ is the mesh size in $x$, $y$ or $z$ directions. The renormalized $J_{AFM}$ is then calculated by considering a fixed exchange coupling energy density as the mesh size varies. The exchange coupling energy in AFM is approximated as the summation of the number of nearest neighbors, which is expressed as

$$J_{AFM}\left\{\frac{[(nx-2)\times(ny-2)\times(nz-2)]\times 2+[(nx-2)\times(ny-2)\times 2\times 2+(nx-2)\times(nz-2)\times 2\times 2+(ny-2)\times(nz-2)\times 2]+[(nx-2)\times 4\times 2+(ny-2)\times 4+(nz-2)\times 4]+8}{\Delta_x^2} + \frac{[(nx-2)\times(ny-2)\times(nz-2)]\times 2+[(nx-2)\times(ny-2)\times 2\times 2+(nx-2)\times(nz-2)\times 2\times 2+(ny-2)\times(nz-2)\times 2\times 2]+[(nx-2)\times 4+(ny-2)\times 4\times 2+(nz-2)\times 4]+8}{\Delta_y^2} + \frac{[(nx-2)\times(ny-2)\times(nz-2)]\times 2+[(nx-2)\times(ny-2)\times 2+(nx-2)\times(nz-2)\times 2\times 2+(ny-2)\times(nz-2)\times 2\times 2]+[(nx-2)\times 4+(ny-2)\times 4+(nz-2)\times 4\times 2]+8}{\Delta_z^2}\right\}$$

where $nx$, $ny$, and $nz$ are the number of grid points in $x$, $y$ or $z$ directions when mesh size varies. The comparison of the switching dynamics of AFM when the mesh size varies from 0.4×0.4×0.4 nm$^3$ to 20×20×2 nm$^3$ is shown in Figure S 2.

To renormalize the interface exchange coupling coefficient ($J_{int}$) between BFO and CoFe, we compare the magnitude of the coercive field of the BFO/CoFe heterojunction devices when the length and width of the device are both 100 nm, and the thickness of BFO and CoFe layers are 30 nm and 2 nm, respectively. The comparison of the coercive field under varying mesh sizes is shown in Figure S 3.

Table 1 Simulation parameter in ferroelectric model[27]

| Variable | Value | Units (SI) |
|---|---|---|
| $P_s$ | 0.8 | $Cm^{-2}$ |
| $\alpha_1$ | 4.9(T-1103)×$10^5$ | $C^{-2}m^2N$ |
| $\alpha_{11}$ | 6×$10^8$ | $C^{-4}m^6N$ |
| $\alpha_{12}$ | -1×$10^6$ | $C^{-4}m^6N$ |
| $\gamma_{FE}$ | 5×$10^{-3}$ | msec/F |
| $K_{Strain}$ | 6×$10^8$ | N/m$^2$ |
| $\epsilon_r$ | 54 | – |
| $E_{ext}$ | 3×$10^8$ | V/m |

Table 2 Simulation parameters in GMR and M-H loop measurement (mesh size is 20×20×2 nm$^3$)

| Variable | Value | Units (SI) |
|---|---|---|
| $M_{s,AFM}$ | 3.6×$10^5$ | A/m |
| $M_{s,FM}$ | 1×$10^6$ | A/m |
| $K_{AFM}$ | -5×$10^5$ | J/m$^3$ |
| $K_{EPI}$ | -5×$10^5$ | J/m$^3$ |
| $A_{AFM}$ | -2.47×$10^{-10}$ | J/m |
| $A_{FM}$ | 1×$10^{-11}$ | J/m |
| $J_{int}$ | 3.2×$10^{-13}$ | J/m |
| $H_{DMI}$ | 1.26×$10^5$ | Oe |
| $\alpha$ | 0.01 | - |
| $\gamma$ | 2.21×$10^5$ | $(A/m)^{-1}s^{-1}$ |
| B | -300~300 | Oe |

*Table 3 Simulation parameters in micromagnetic model (mesh size is 5×5×1 nm³)*

| Variable | Value | Units (SI) |
|---|---|---|
| $M_{s,AFM}$ | $3.6 \times 10^5$ | A/m |
| $M_{s,FM}$ | $1 \times 10^6$ | A/m |
| $K_{AFM}$ | $-2 \times 10^5$ | J/m³ |
| $K_{EPI}$ | $-2 \times 10^5$ | J/m³ |
| $A_{AFM}$ | $-6.07 \times 10^{-11}$ | J/m |
| $A_{FM}$ | $1 \times 10^{-11}$ | J/m |
| $J_{int}$ | $5 \times 10^{-13}$ | J/m |
| $H_{DMI}$ | $1.45 \times 10^5$ | Oe |
| $\alpha$ | 0.01 | - |
| $\gamma$ | $2.21 \times 10^5$ | $(A/m)^{-1} s^{-1}$ |

# Supplementary materials

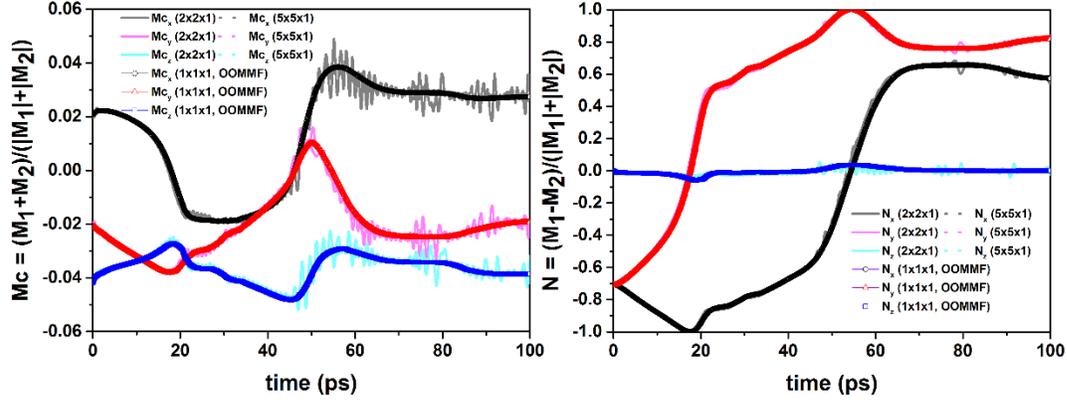

*Figure S 1 Comparison of the switching curves of (a) the weak magnetization (Mc) and (b) the Neel vector (N) using OOMMF versus MATLAB. In the MATLAB, the exchange coupling in the antiferromagnet is expressed as $H_{exch}^{ij} = \frac{2A}{\mu_0 M_s a^2}\left(\alpha_i^x, \alpha_i^y, \alpha_i^z\right)$ which is independent of the mesh size.*

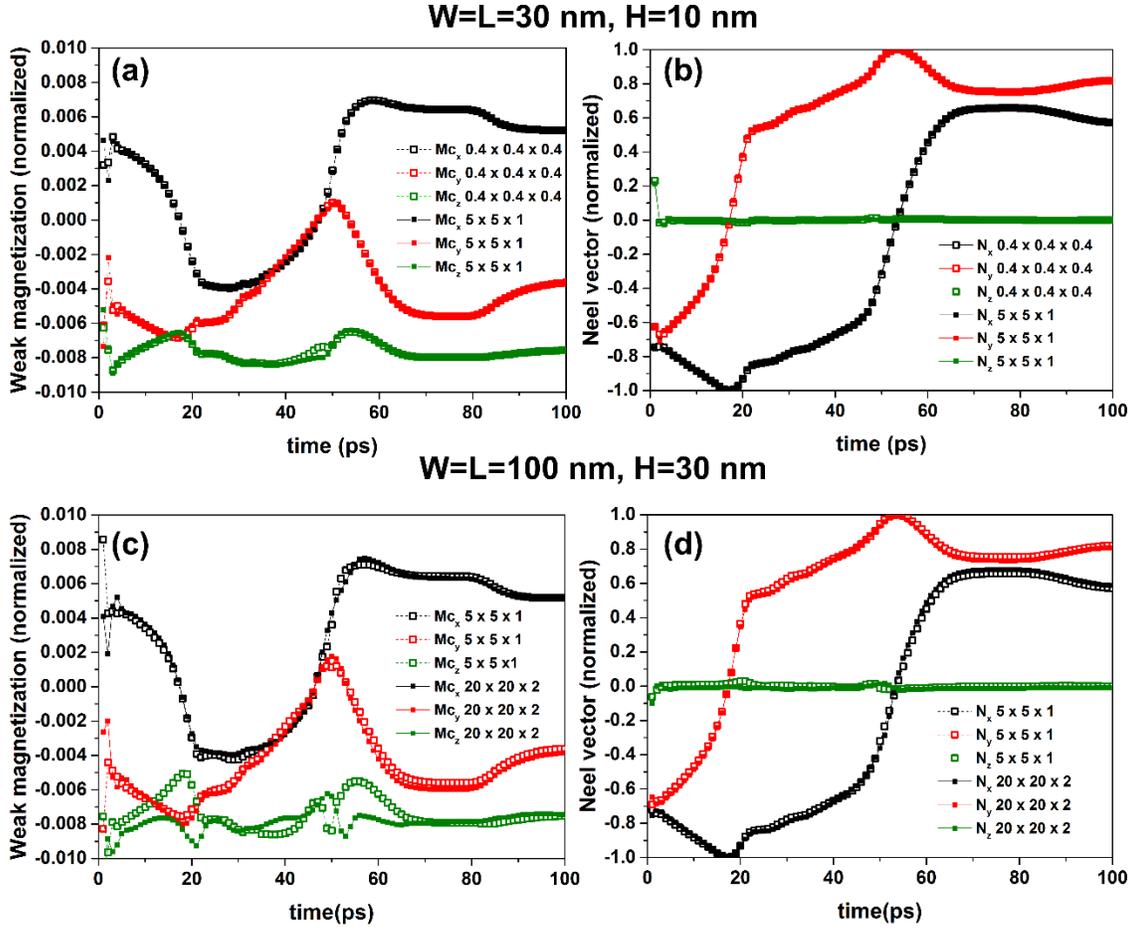

*Figure S 2 The switching curves of (a) the weak magnetization (Mc) and (b) the Neel vector (N) of BFO under renormalized exchange stiffness constant ($J_{AFM}$) from 0.4×0.4×0.4 to 5×5×1 when width and length are equal to 30 nm, and height is 10 nm thick. The switching curves of (c) N and (d) Mc of BFO under renormalized $J_{AFM}$ from 5×5×1 to 20×20×2 when width and length are equal to 100 nm, and height is 30 nm thick.*

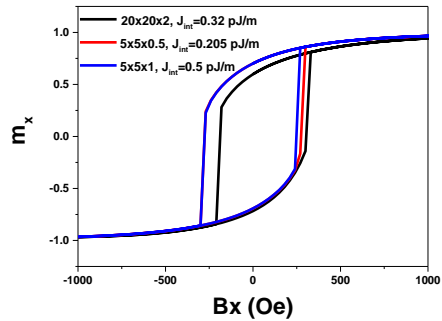

*Figure S 3 Comparison of the magnetic hysteresis loop under varying mesh size (20×20×2, 5×5×0.5, 5×5×1) with constant magnetic coercive field.*